\documentclass[11pt,a4paper]{article}
\usepackage{jinstpub}
\usepackage{verbatim}
\usepackage{graphicx}
\usepackage{subcaption}
\usepackage[sort&compress]{natbib}
\def\ECO{\rm C_3F_4H_2}
\def\FREON{\rm C_2F_4H_2}
\def\SF6{\rm SF_6}
\def\CO2{\rm CO_2}
\title{MRPC with eco-friendly gas}
\author[a,b,1]{Y.W. Baek,\note{Corresponding author.}}
\author[a,c]{D.W. Kim,}
\author[a,c]{W.S. Park,}
\author[a,b,d]{M.C.S. Williams,}
\author[c,e]{and R. Zuyeuski,}

\affiliation[a]{Gangneung-Wonju National University,\\
Gangneung, South Korea}
\affiliation[b]{CERN,\\
Geneva, Switzerland}
\affiliation[c]{ICSC World Laboratory,\\
Geneva, Switzerland}
\affiliation[d]{INFN and Dipartimento di Fisica e Astronomia,\\
Universit di Bologna, Italy}
\affiliation[e]{Museo Storico della Fisica e Centro Studi e Ricerche E. Fermi,\\
Roma, Italy}
\emailAdd{yong.wook.baek@cern.ch}

\abstract{The Multigap Resistive Plate Chambers (MRPC) are used as a timing detector in several particle physics and cosmic ray experiments.  The gas mixture of MRPC at current experiments is a mixture containing $\FREON$ and in some cases $\SF6$.  $\FREON$ and $\SF6$ have a Global Warming Potential (GWP) of 1430 and 23900 respectively, therefore they are classified as greenhouse gases.
The studies to reduce the amount of emission of the greenhouse gas in high energy experiments are underway; the present contribution has been performed as part of this effort. The results have been obtained from the beam test of a small MRPC which has 6 gaps of 220 $\mu$m and an sensitive area of 20 $\times$ 20 $\rm cm^2$. It has been operated with the ecological HFO-1234ze gas ($\ECO$), and with the  $\FREON$/$\SF6$ mixture. We have found that the ecological gas can substitute for the $\FREON$-based gas mixture without significantly compromising the current level of performance. }

\begin{document}
\maketitle

\section{Introduction}
The Resistive Plate Chamber (RPC) is a gaseous detector that can cover a large area; it has a low cost of construction and a high quality of performance. Most RPC detectors have been operating with the $\FREON$-based gas mixture which has very high Global Warming Potential  (GWP) value. In many experiments, as an expedient to reduce amount of gas emitted into the atmosphere, a closed loop gas system has been introduced. However, the construction cost is not negligible and the problem of gas leaks still remains. A better alternative is to use an ecological gas.  Studies to find an ecological gas mixture have been performed by some groups~\cite{eco:helium, eco:RPC}; this study is carried out as part of this program. 

The present contribution describes the result obtained with Multigap RPC operating with a gas mixture based on the ecological gas (HFO-1234ze: $\ECO$, GWP <7).  As a comparison, the results obtained with the gas mixture that includes R134a ($\FREON$) is also presented.

\section{Multigap Resistive Plate Chamber}
The test Multigap Resistive Plate Chamber (MRPC) has an active area of 20 $\times$ 20 $\rm cm^2$. It has  24  pickup strips of 0.7 $\times$ 20.5 $\rm cm^2$ and 6 gaps of 220 $\mu$m with a glass thickness of 280 $\mu$m. The time resolution is 80 ps with the $\FREON$/$\SF6$ gas mixture~\cite{FORSTER2016182}.  The ultrafast NINO amplifier-discriminator~\cite{Anghinolfi:818554} that has been developed for ALICE TOF detector at LHC~\cite{Aamodt:2008zz} is used.  The NINO card receives differential input from MRPC and produces LVDS output signal; the width of the output signal depends on the input charge. The time slewing correction has  been performed using the measurement of the width.

\begin{figure}[h]
\centering
\includegraphics[width=0.6\textwidth]{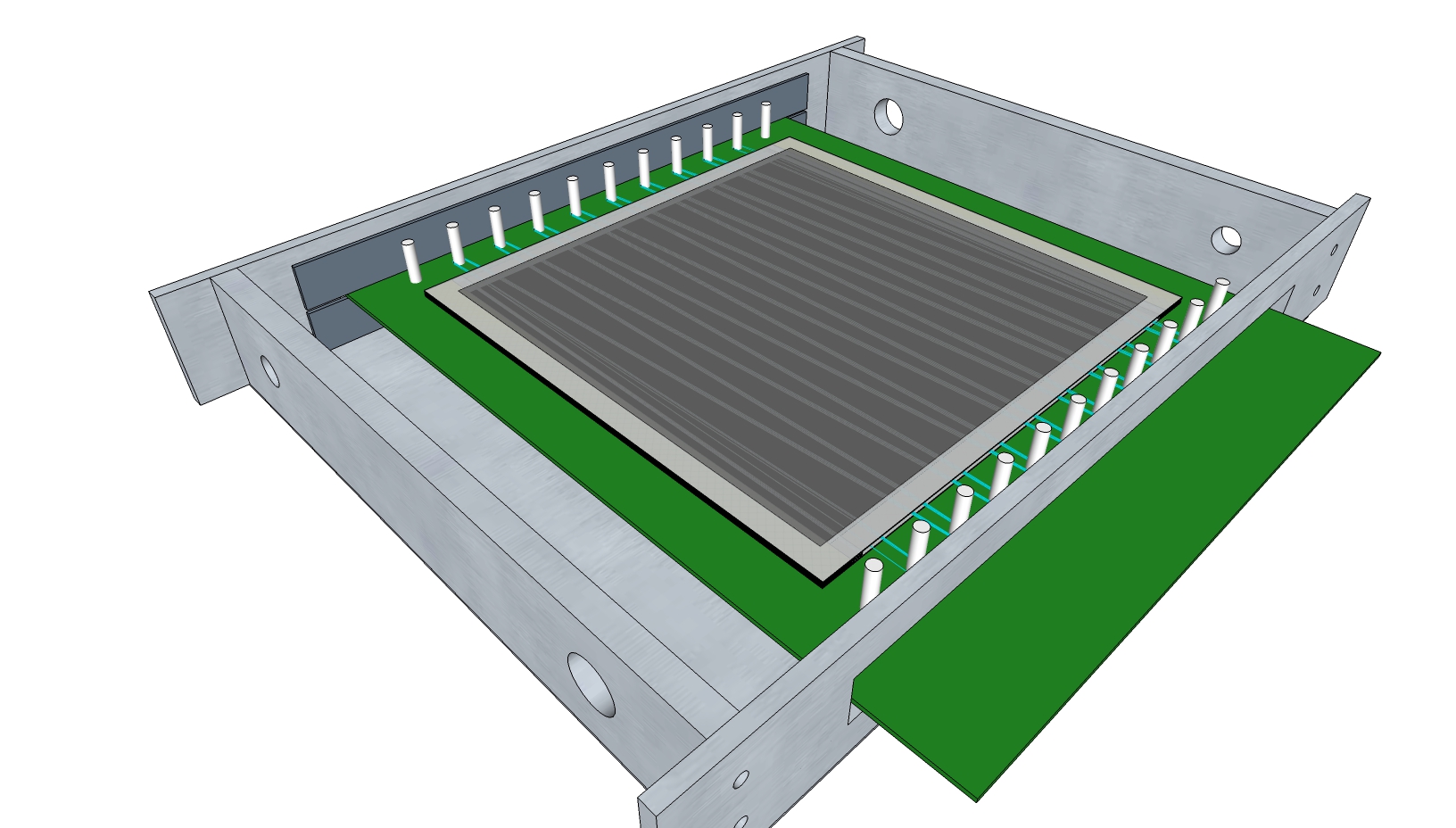}
\caption{The MRPC used for the beam test. It has 20 $\times$ 20 $\rm cm^2$ active area with 6 gaps of 220 $\mu$m.}
\label{smallMRPC}
\end{figure}

Various gas mixtures have been tested, the tetrafluoroethane-based mixture (95\% $\FREON$ and 5\% $\SF6$) and the tetrafluoropropene-based mixtures ($\ECO$ with $\CO2$ or $\SF6$) with different gas component ratios.    

For the readout of the signals, NINO cards were mounted at both ends of the strips. One NINO card has three NINO chips; it can readout signals from 24 strips.  The threshold was set at 160 mv at the NINO threshold input (this requires a signal of $\sim$ 40 fC to fire the NINO). The time-stamp of the leading and trailing edge of the LVDS output signal from the NINO were given by two CAEN V1290A TDCs that have a time resolution of 30 ps.

\section{Experimental setup}
\begin{figure}[th]
\centering
\includegraphics[width=0.8\textwidth]{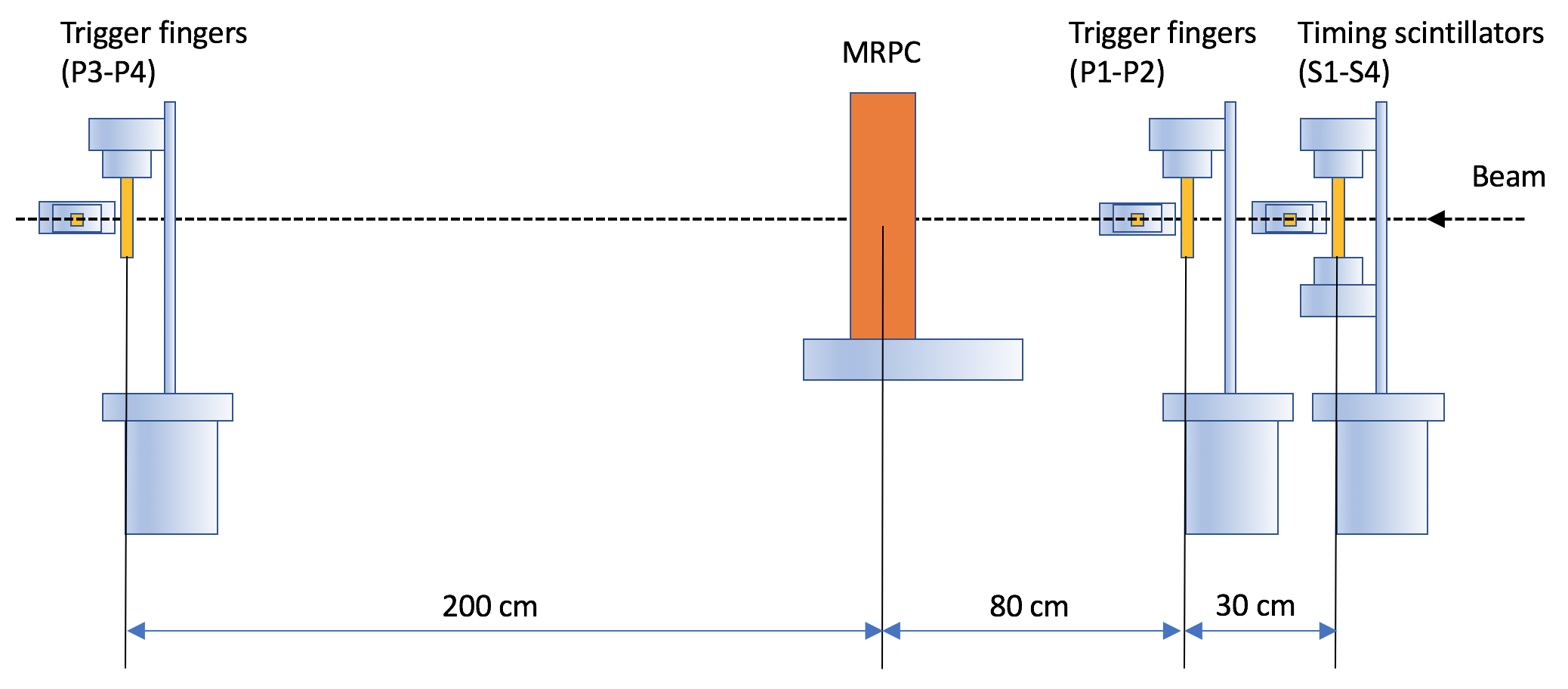}
\caption{Setup at T10 beam test facility}
\label{T10setup}
\end{figure}

The MRPC test with the ecological gas has been carried out at T10, a test beam in the CERN East hall, that provides a beam of pions with a momentum up to 6 GeV/c.  The setup at T10 is shown in figure~\ref{T10setup}. 
Two sets of trigger finger scintillators and a pair of timing scintillators are used select through going particles.  The trigger finger scintillators have active areas of $1 \times 1 \rm \;cm^2$  for P1$-$P2 and $2 \times 2 \rm \;cm^2$ for P3$-$P4. The coincidence of these scintillators is used as the trigger.   Two fast scintillator bars ($2 \times 2 \times 10 \rm \;cm^3$) each equipped with two photomultipliers, S1$-$S4, provide an accurate time reference.

\section{Data analysis and results}
During the first test beam period, the data is taken at an instantaneous flux of 1.3\,kHz/$\rm cm^2$.  During the second period the performance of the MRPC was measured with different particle flux.  
To have a good reference time we selected events that should be within $\pm 3\sigma$ from the mean of the time difference of two scintillators; (S1+S2)/2 - (S3+S4)/2. From the obtained $\sigma$, the precision of the mean time used as reference time, (S1+S2+S3+S4)/4, can be estimated.  The resolution of this reference time is measured to be 47 ps and it has been subtracted in quadrature when estimating the time resolution of the MRPC.

\subsection{Efficiency and rate capability}
\begin{figure}[h]
\centering\includegraphics[width=0.6\textwidth]{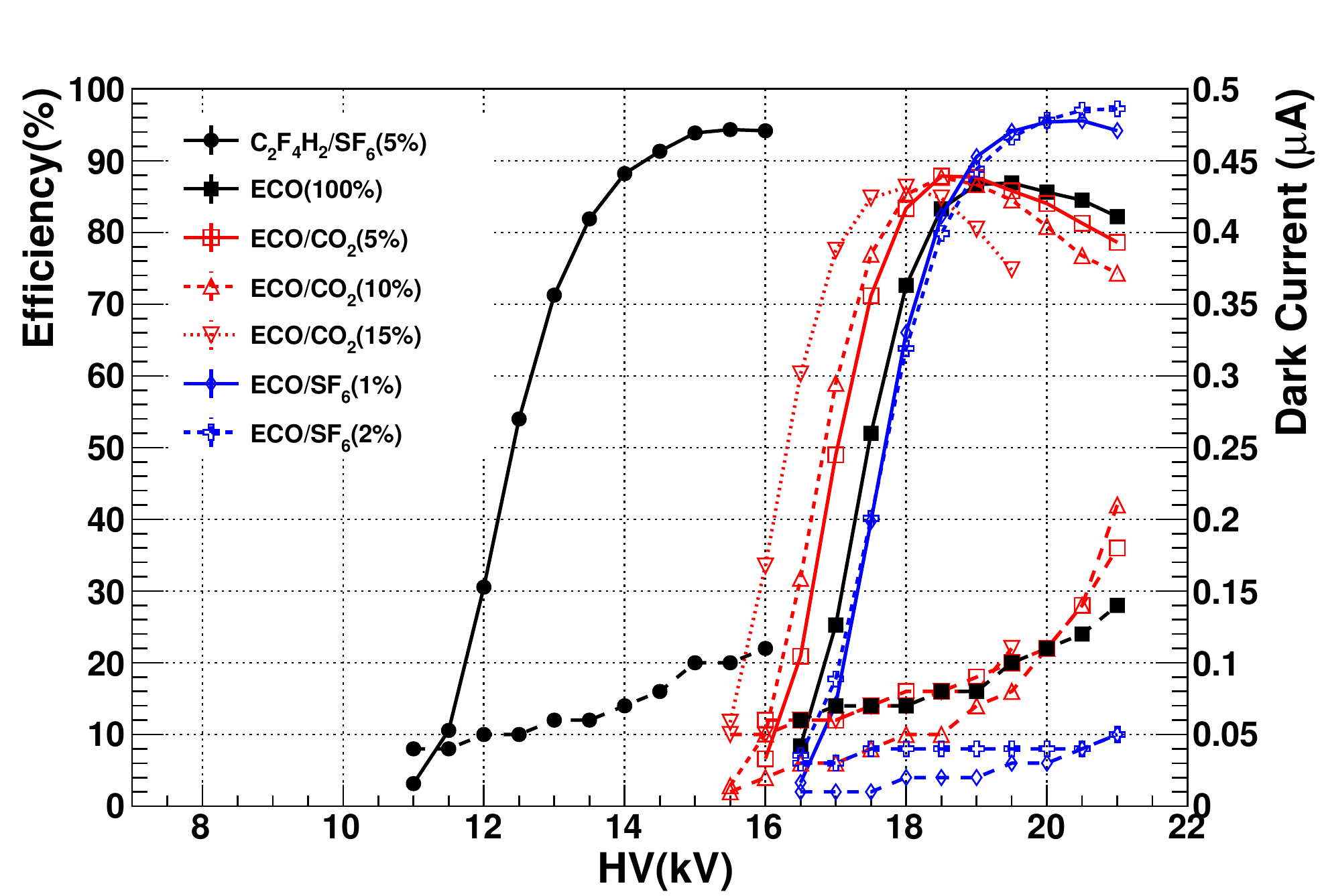}
\caption{Efficiencies measured with various gas mixtures at 1.3\,kHz/$\rm cm^2$ of particle flux. Dark currents are shown with dashed lines.}
\label{efficiency}
\end{figure}

The chamber efficiency is defined as the ratio between number of events detected by the MRPC and the number of triggered events.  The measured efficiencies for various gas mixtures are shown in figure~\ref{efficiency}. 
\begin{figure}[!h]
\centering\includegraphics[width=0.6\textwidth]{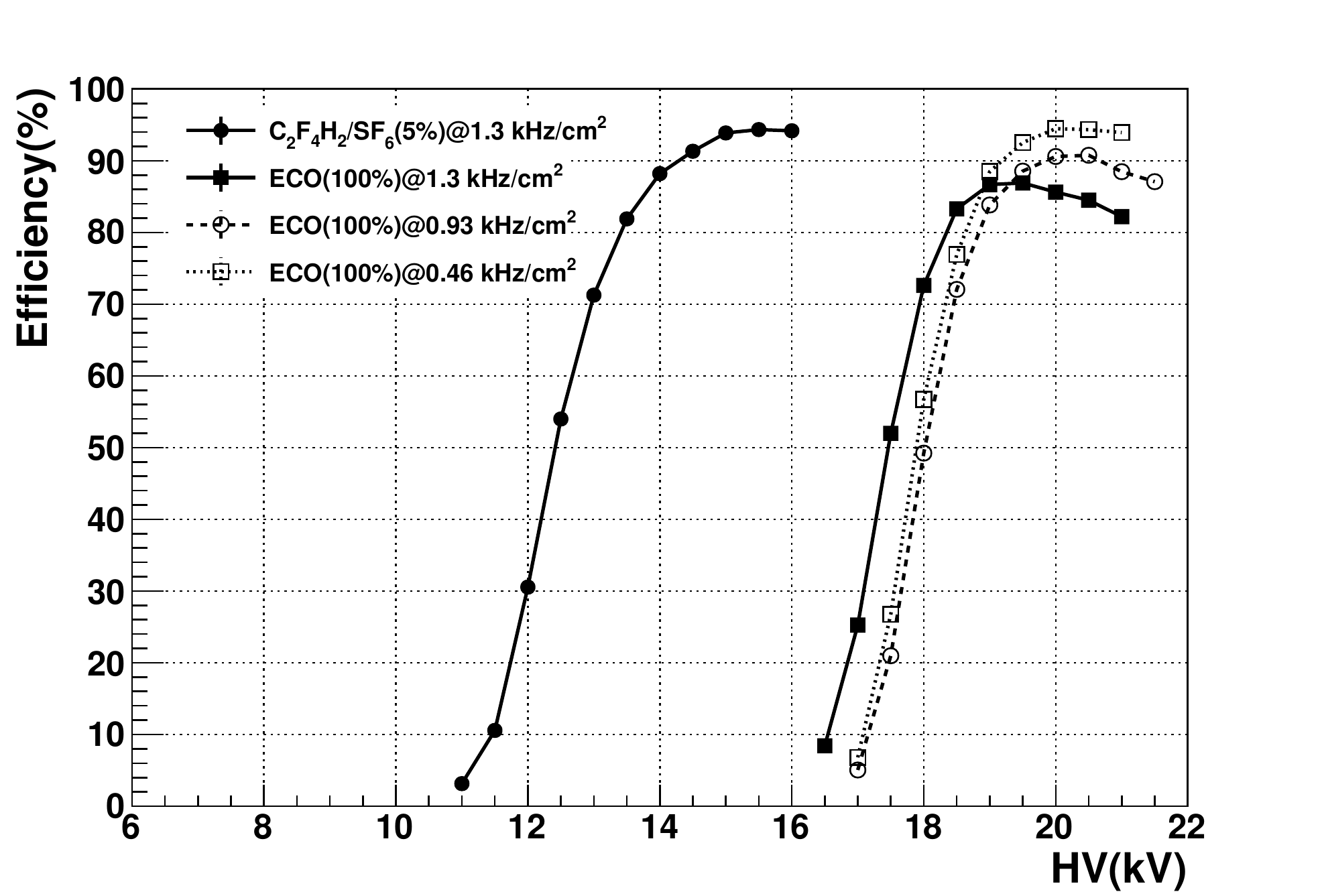}
\caption{Efficiencies for pure ecological gas at different counting rates.}
\label{fig:rate}
\end{figure}
In the case of the ecological gas mixtures, the operating voltages need to be higher by 4\,kV than those with the $\FREON$ based gas mixture.  The efficiencies are below 90\%  when ecological gas mixtures are used with and without $\CO2$.  The effect of adding $\CO2$ is to shift the efficiency plateau to lower voltage.  
When 15\% of $\CO2$ is used, this shift amounts to 1.2 kV. However, we observe that the efficiency plateau became narrower in this case. The fast drop of the efficiency at higher voltages is correlated to the higher values of the dark current, which implies an increase of the streamer production. If a small amount of $\SF6$ is added to the ecological gas, we obtain again high efficiency and low dark current.

During the second test beam period, the same chamber has been tested at different particle flux. 
The counting rate is monitored by fast timing scintillator bars (S1$-$S4) with 2 $\times$ 2 cm$^2$ active area. The time of a spill is 350 ms and the coincidence of four fast scintillator bars (S1$-$S4 with active area, 2 $\times$ 2 \rm cm$^2$) is divided by the spill time and the 4 cm$^2$ active area.
The results for an instantaneous flux of 930\,Hz/cm$^2$ and  460\,Hz/cm$^2$, are shown in figure~\ref{fig:rate}.  Maximum efficiency values of 95\% and 90\% were obtained for the rate of 460 Hz/cm$^2$ and 930 Hz/cm$^2$, respectively.  

\subsection{Time resolution}
\begin{figure}[!b]
\centering\includegraphics[width=0.6\textwidth]{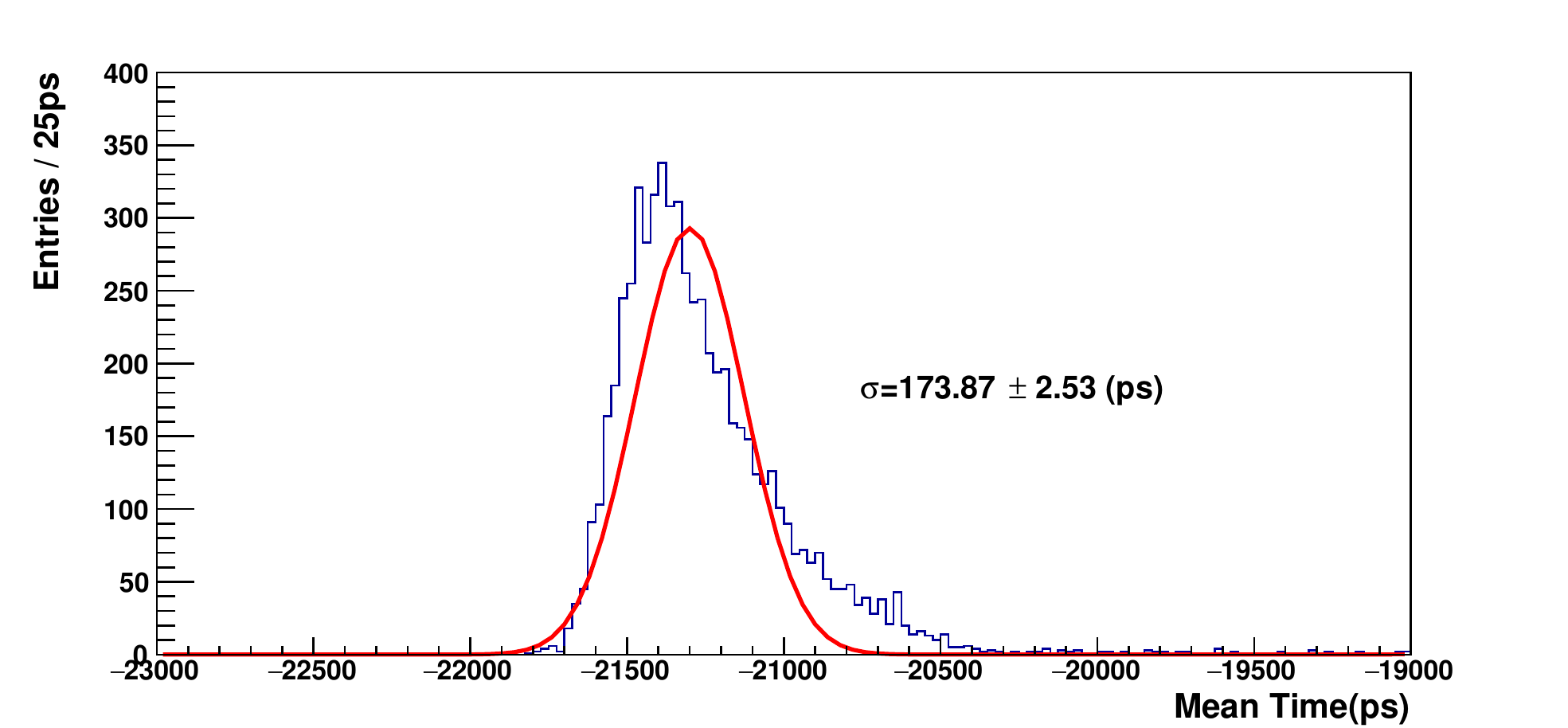}
\caption{Distribution of mean times}
\label{fig:rawMTResol}
\end{figure}
\begin{figure}[!b]
\centering\includegraphics[width=0.6\textwidth]{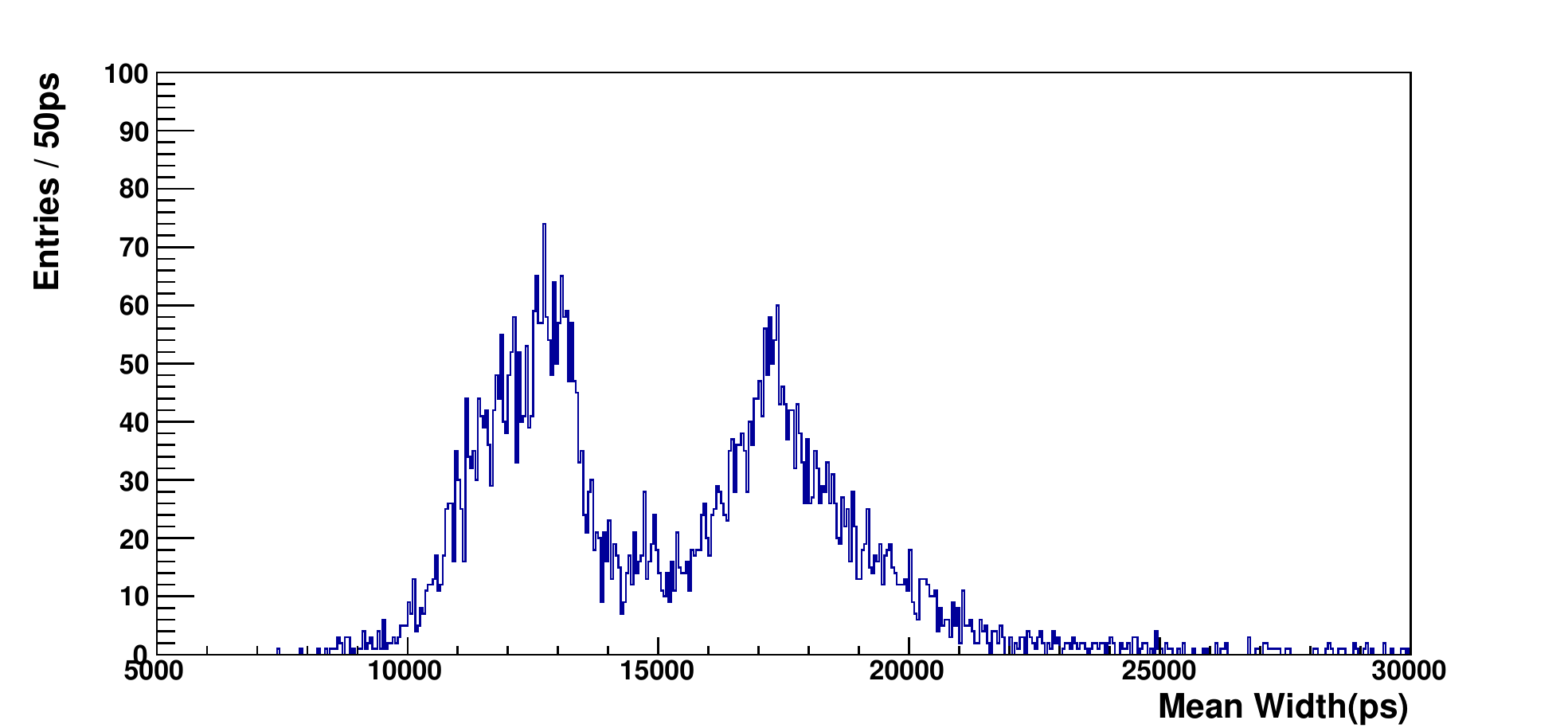}
\caption{Distribution of mean pulse widths.}
\label{fig:meanwidth}
\end{figure}
\begin{figure}[!h]
\centering\includegraphics[width=0.6\textwidth]{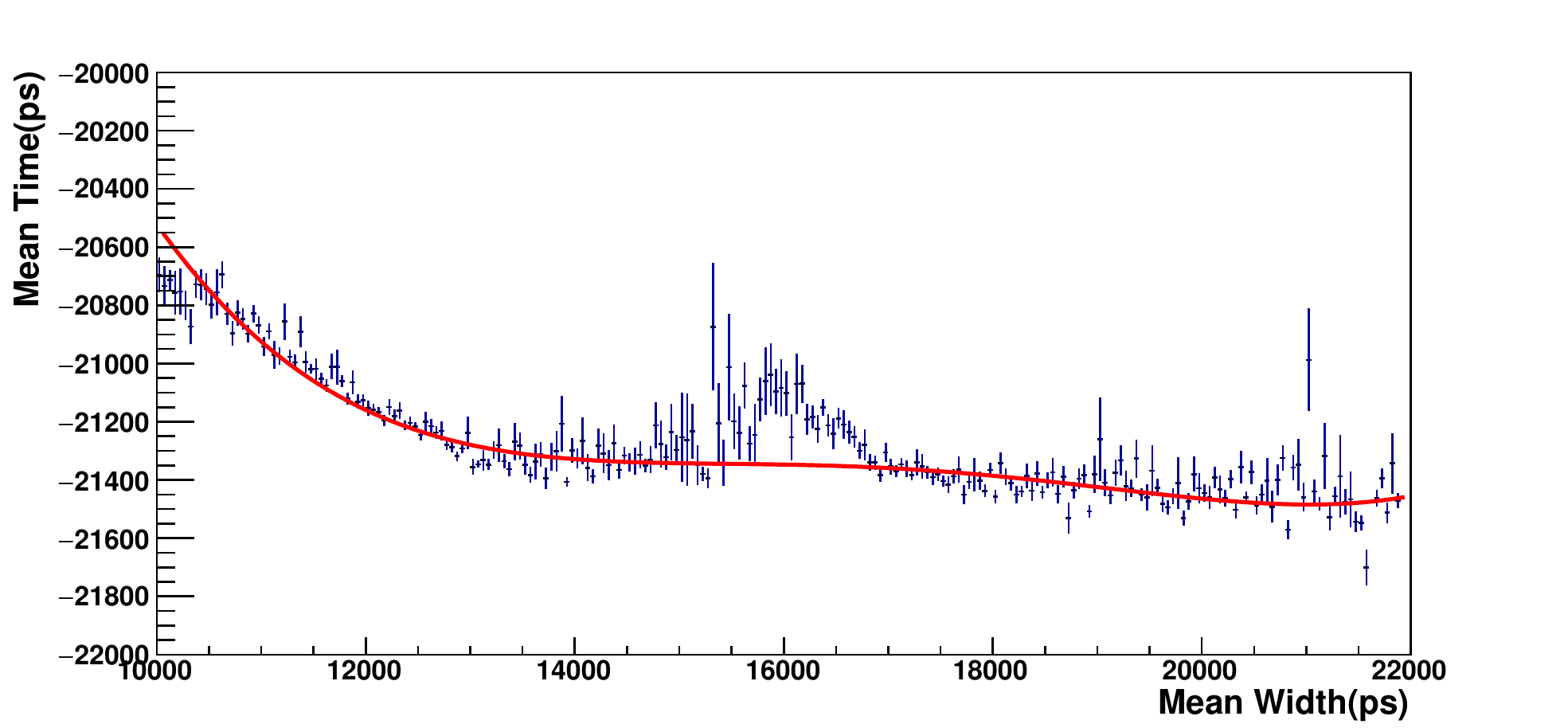}
\caption{Profile of mean time vs. mean pulse width and a fourth order polynomial function for the time-slewing correction.}
\label{fig:TA}
\end{figure}
\begin{figure}[!h]
\centering\includegraphics[width=0.6\textwidth]{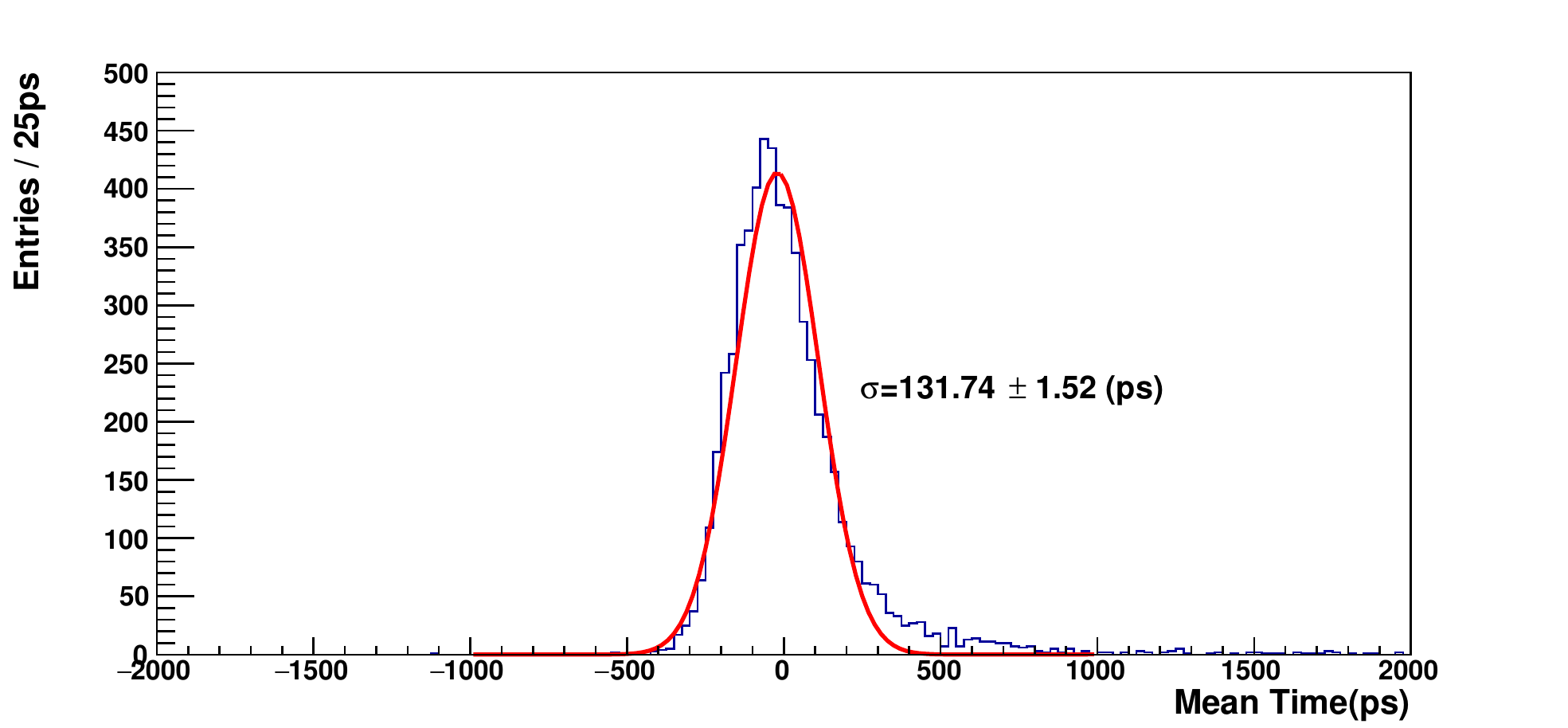}
\caption{Time-slewing corrected mean time distribution.}
\label{fig:corr}
\end{figure}

The signal from the MRPC is readout at both ends of the strips. As the time of a given event, we used the mean time $\rm t_{mean} =  ((t_{end1} - t_{ref}) + (t_{end2} - t_{ref}))/2$, where $\rm t_{end1}$ and $\rm t_{end2}$ are the time of the leading edges of the signal measured by the NINO and $\rm t_{ref}$ is the reference time from the timing scintillators. The distribution of the mean time is shown in figure~\ref{fig:rawMTResol}, for the data obtained at 19 kV and with 100\% of $\ECO$.

In figure~\ref{fig:meanwidth}, the distribution of the mean of the pulse widths measured at both ends of the strips is plotted. Using a fixed threshold introduces a time shift (slewing) depending on the signal size, which requires a time-slewing correction. A fourth order polynomial $f$ as a function of mean width was fitted to the profile distribution as in figure~\ref{fig:TA}. The distribution of the corrected time,  $\rm t_{corrected} = t_{\rm mean} - {\it f}$, is presented in figure~\ref{fig:corr} with its gaussian fit result. The actual time resolution of the  MRPC is obtained by subtraction the jitter of timing scintillators ($\sim 47$ ps) in quadrature.

\begin{figure}[h]
\centering\includegraphics[width=0.6\textwidth]{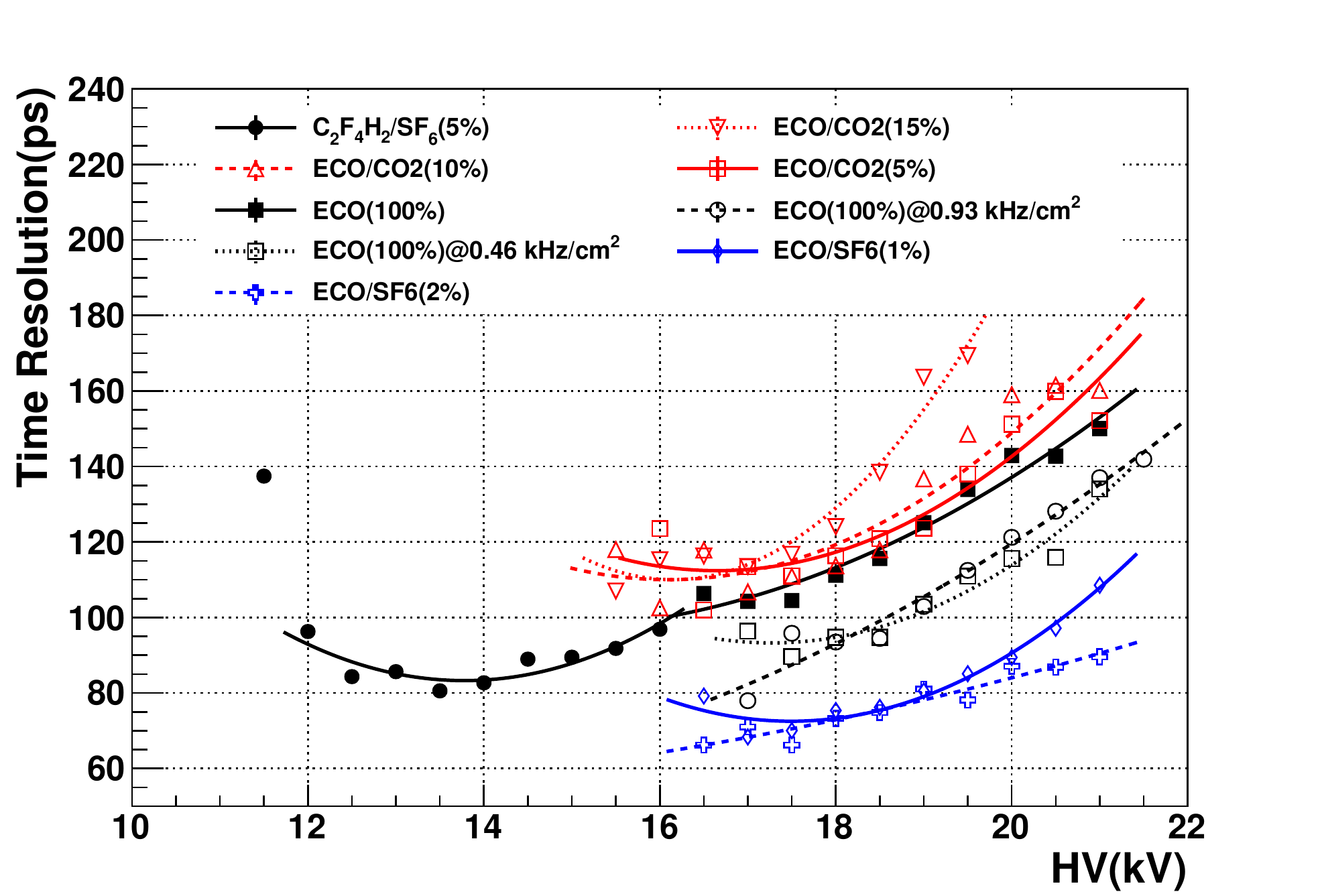}
\caption{Mean time resolution.}
\label{fig:MTresolution}
\end{figure}

The time resolutions measured with different gas mixtures are shown in figure~\ref{fig:MTresolution}.  With the $\FREON$/$\SF6$ gas mixture, we obtained 90 ps at 15 kV. When pure ecological gas is used, the resolution at 19 kV is 120 ps. Adding $\CO2$ to pure ecological gas shifts the operation voltage to a lower value while keeping the time resolution almost the same as with the pure ecological gas. Reducing the beam intensity improved the time resolution slightly; 110 ps at 20 kV.
\begin{figure}[h]
\centering\includegraphics[width=0.6\textwidth]{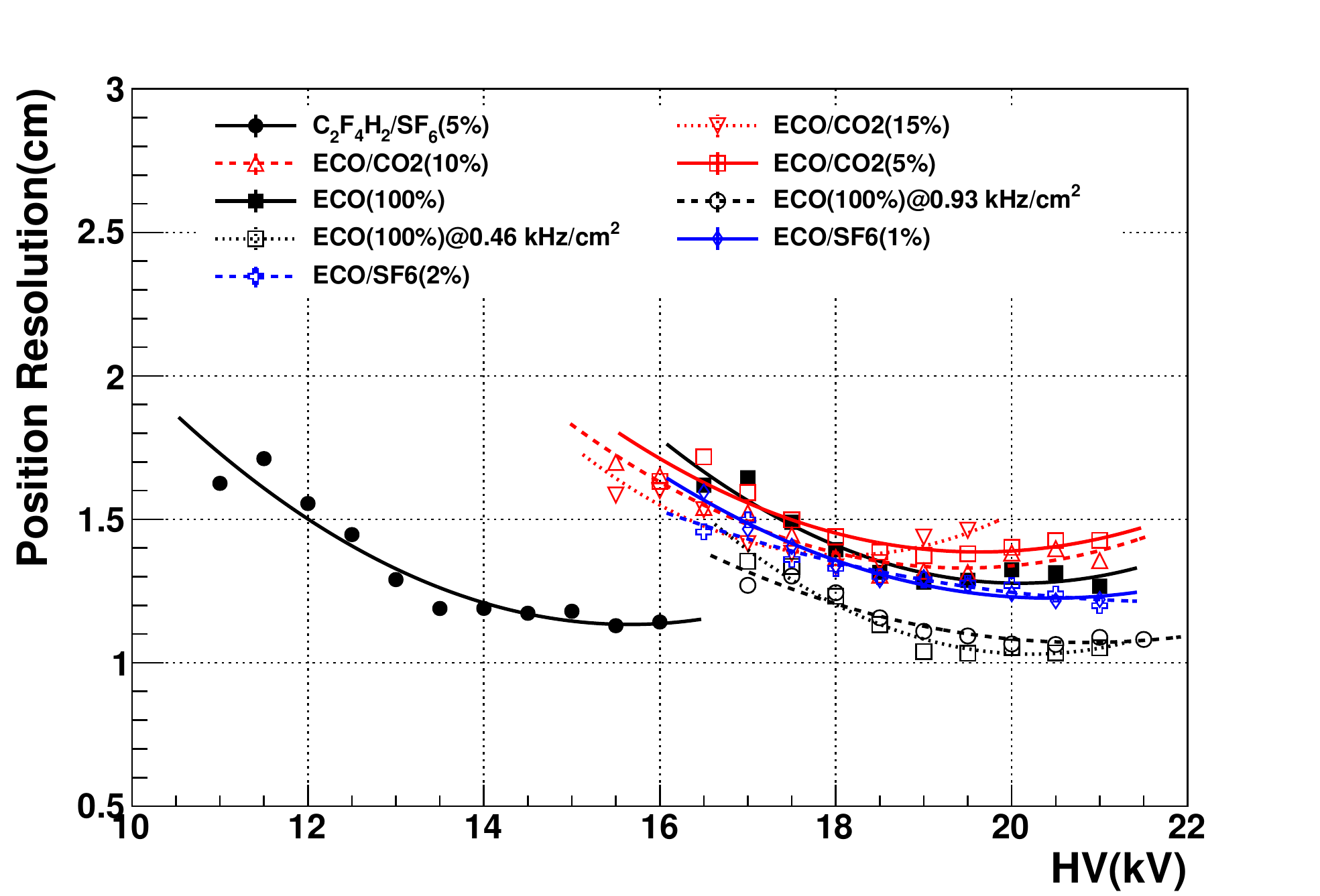}
\caption{Position resolution.}
\label{fig:positionAll}
\end{figure}
\begin{figure}[h]
\centering\includegraphics[width=0.6\textwidth]{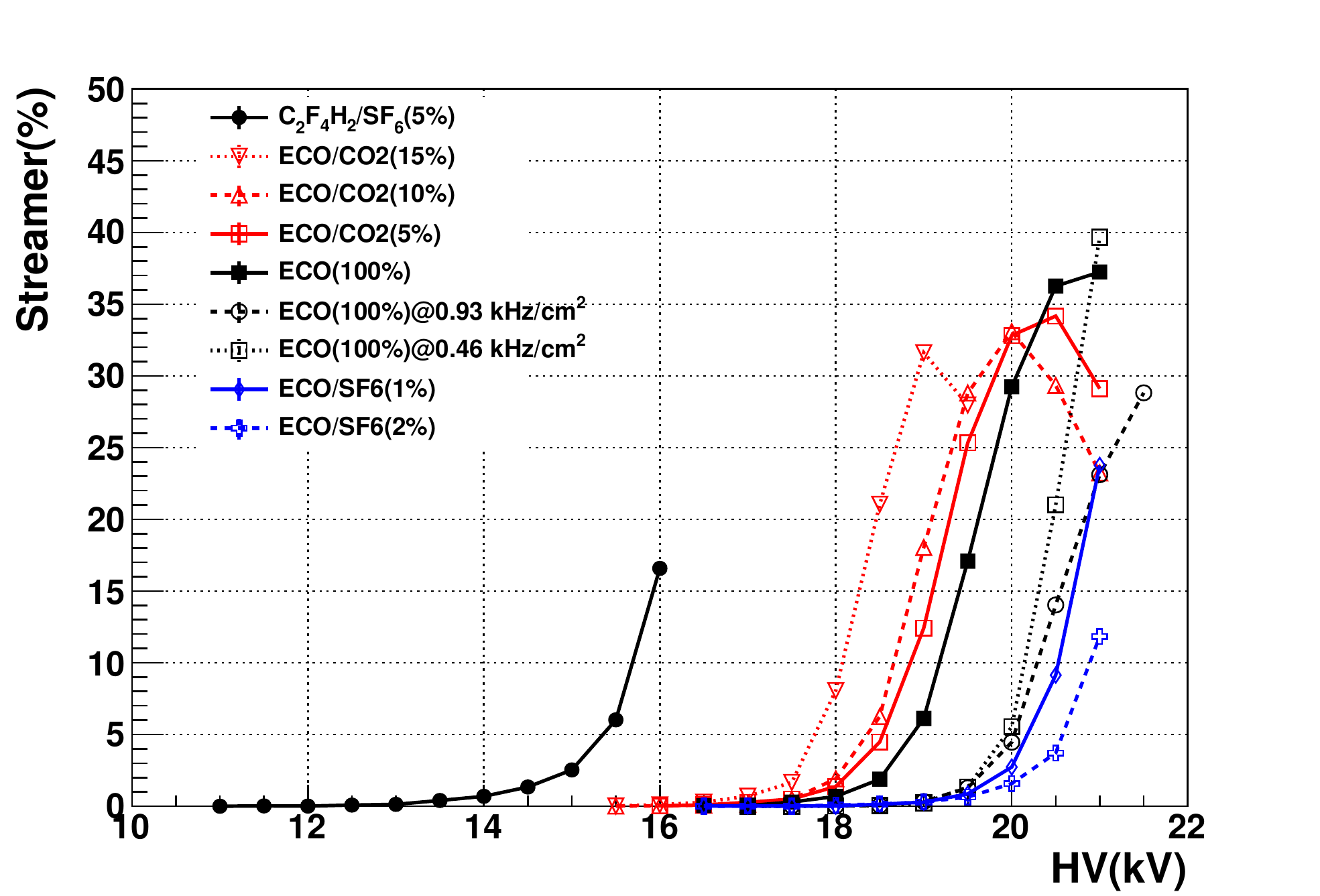}
\caption{Streamer Rate}
\label{fig:streamers}
\end{figure}

\subsection{Position resolution}
The difference in time values measured at both ends of the strip gives information on the hit position. The precision of the position measurement can be estimated by the distribution of this time difference. 
Here two thirds of the speed of light is taken as the speed of signal on the strip. The measured position resolutions for all gas mixtures are shown in figure~\ref{fig:positionAll}. Depending on the gas mixture types and the rate, they vary from 1 cm to 1.3 cm at 20 kV.

\subsection{Streamer rate}
The streamer rate and its size can have effects on the detector performance, especially on the rate capability and the life time of the chamber.  We defined the event as streamer if more than 5 consecutive strips gave the signal over the threshold of NINO on both sides.  
The operation with ecological gas mixtures shows a higher streamer rate than with the $\FREON$/$\SF6$ gas mixture as shown in figure~\ref{fig:streamers}.
Adding $\CO2$ has an effect of reducing the operating voltage, but it does not suppress the streamer production at the knee of the efficiency plateau significantly unlike the case of adding $\SF6$.

\section{Conclusion}
The performance of a MRPC with 6 gaps of 220 $\mu$m has been measured using the ecological gas (HFO-1234ze:$\ECO$, GWP $<$ 7). A shift of 4 $\sim$ 5 kV to higher values of the applied voltage has been observed in efficiency curve when pure ecological gas was used, and an efficiency of 95\% has been obtained for a low incident particle rate of 460 Hz/cm$^2$. The time resolution has been measured to be 120 ps, compared with 90 ps when using the $\FREON$/$\SF6$ gas mixture. The effect of adding the $\CO2$ has also been studied. Although this gas does not suppress the streamers (as does $\SF6$), it shifts the efficiency curve to a lower value of the high voltage by 1 kV. 

\section*{Acknowedgements}
This work has been supported by the Korea-EU cooperation program of National Research Foundation of Korea, Grant Agreement 2014K1A3A7A03075053. The results presented here were obtained at the T10 test beam in the east hall at CERN. The authors acknowledge the support received by the operators of the PS.





\bibliographystyle{plainnat}
\bibliography{references}







\end{document}